\documentclass[useAMS,usenatbib]{mn2e}
\usepackage{graphicx}
\usepackage[a4paper]{hyperref}
\usepackage{amssymb}

\title[Radio-Optical Behaviour and SED of the Blazar GC
0109+224]{Radio-optical flux behavior and spectral energy
distribution of the intermediate blazar GC 0109+224}
\author[Ciprini, Tosti, Ter\"{a}sranta, Aller]
    {Stefano Ciprini$^{1,2}$\thanks{offprints: stefano.ciprini@pg.infn.it},
    Gino Tosti$^{1,2}$, Harri Ter\"{a}sranta$^3$, and Hugh D. Aller$^4$
  \\
  $^1$Physics Department and Astronomical Observatory, University
of Perugia, via Pascoli, 06123 Perugia, Italy\\
  $^2$INFN Perugia Section, via Pascoli, 06123 Perugia, Italy\\
  $^3$Mets\"{a}hovi Radio Observatory, Helsinki University of Technology, 02540, Kylm\"{a}l\"{a},
  Finland\\
  $^4$Department of Astronomy, Dennison Bldg., University of Michigan, Ann Arbor, MI
48109, USA
  }
\date{submitted to MNRAS}

\pagerange{\pageref{firstpage}--\pageref{lastpage}} \pubyear{2003}

\begin{document}

\label{firstpage}

\maketitle

\begin{abstract}
About twenty years of radio observations in five bands (from 4.8
to 37 GHz) of the BL Lac object GC 0109+224 (S2 0109+22, RGB
J0112+227), are presented and analysed together with the optical
data. Over the past ten years this blazar has exhibited enhanced
activity. There is only weak correlation between radio and optical
flares delays, usually protracted on longer timescales in the
radio with respect to the optical. In some cases no radio flare
counterpart was observed for the optical outbursts. The radio
variability, characterised by peaks superposition, shows hints of
some characteristic timescales (around the 3-4 years), and a
fluctuation mode between the flickering and the shot noise. The
reconstructed spectral energy distribution, poorly monitored at
high energies, is preliminarily parameterised with a
synchrotron-self Compton description. The smooth synchrotron
continuum, peaked in the near-IR-optical bands, strengthens the
hypothesis that this source could be an intermediate blazar.
Moreover the intense flux in millimetre bands, and the optical and
X-ray brightness, might suggest a possible detectable gamma-ray
emission.
\end{abstract}


\section{Introduction}\label{sec:intro}
The compact radio--loud source GC 0109+224 (S2 0109+22, TXS
0109+224, RX J0112.0+2244, EF B0109+2228, RGB J0112+227),
belonging to the Green Bank radio survey list C, is a synchrotron
source known for more than thirty years \citep{pauliny72} and was
optically identified few years later \citep{owen77}. It is a
strong radio-millimetre active galactic nucleus (AGN), with
variable flux, polarisation degree and position angle, showing a
flat average spectrum much like the classical BL Lac objects.
Unfortunately it is poorly studied beyond optical frequencies. The
milliarcsecond (pc) scale of GC 0109+224 reveals a compact core,
and a secondary component with no additional diffuse emission, and
less luminous and/or beamed than the 1-Jy sample sources
\citep{bondi01,fey00}. The kpc scale shows a faint one--sided
collimated radio jet, about 2 arcsec long \citep{wilkinson98},
largely misaligned with the pc-scale inner region, as is
frequently observed in high-luminosity and low-energy peaked BL
Lac objects (LBLs, with the peak in infrared bands). This source
is a member of the 200-mJy catalog \citep{marcha96}, a sample
which seems to fill the gap between the high energy peaked BL Lacs
(HBLs, with the peak in the UV, soft-X bands) and the LBLs, as
expected by some current blazar unification pictures.
\par The historical optical light curve shows a behavior intermediate between
the larger--amplitude variable BL Lacs, and the smaller--amplitude
variable flat spectrum radio quasar (FSRQs). Increased flare
activity (rapid optical variations, peak superposition, rapid flux
drops), is clearly shown after 1994 in the optical bands, thanks
to the increased data sampling of the Perugia University
Observatory, which has also recorded the larger and faster flares
\citep{ciprini03a}. During seven years GC 0109+224 showed six main
optical outbursts (weeks/month scale), and a variability mode
placed between the flickering and the shot noise.

Rather achromatic long term optical behaviour , is accompanied by
isolated outbursts with loop--like hysteresis of the spectral
index, indicating that rapid optical variability is dominated by
non-thermal cooling of a single electron population
\citep{ciprini03a}. Both the optical flux and the polarization are
known to be variable on different timescales, including intraday
variability \citep{sitko85,mead90,valtaoja91}.
%
\begin{figure*}
\begin{tabular}{c}\hskip -0.5cm
{\resizebox{\hsize}{!}{\includegraphics[angle=90]{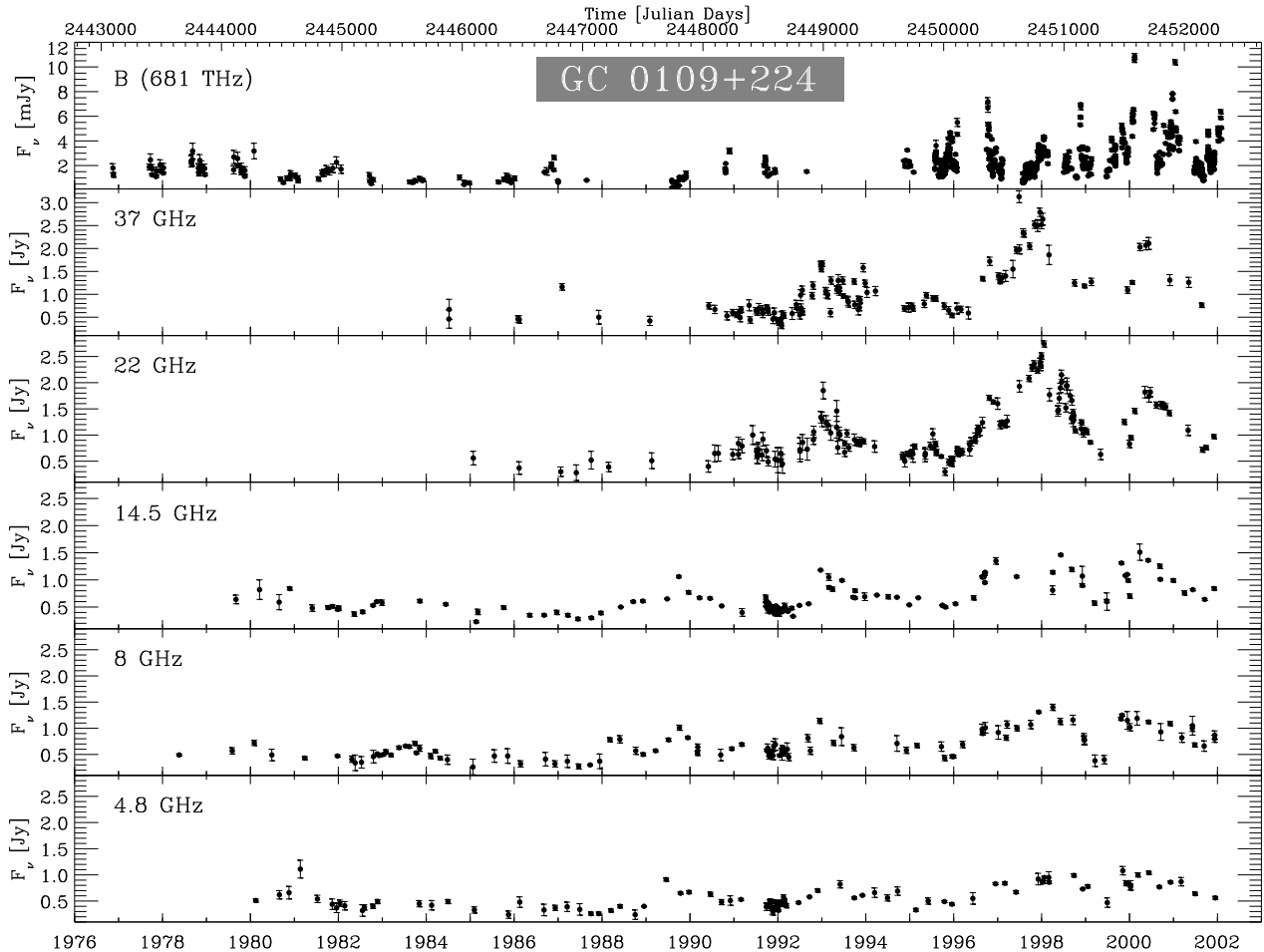}}}\\
\end{tabular}
\vskip -0.5 true cm \caption{The complete radio--optical density
flux light curves of GC 0109+224. The optical observations are
historical, and Perugia Obs. (Italy) data (the post-1994 data
points), reported in the original or extrapolated Johnson B band
\citep[see for details ][]{ciprini03a}. The high sampled data at
37 and 22 GHz are from Mets\"ahovi (Finland), and the 14.5 GHz, 8
GHz, 4.8 GHz data are from the long term monitoring of the UMRAO,
(USA). This source clearly exhibit a higher mean--flux level, and
an enhanced activity in the radio--mm after 1992, and in the
optical after 1994, even if the optical sampling before this year
is insufficient to record faster (and usually larger) flares.}
\label{fig:clucetutte}
\end{figure*}
%
The relatively strong variable degree (up to 30\%) and direction
of the optical linear polarisation, is one of the most noticeable
characteristics of this object \citep{takalo91,valtaoja93}, even
if there is no clear correlation between the flux level and
polarisation degree. The host galaxy of GC 0109+224 is still
unresolved \citep{falomo96,wright98}, but a lower limit to the
redshift $z \ge 0.4$ is suggested \citep{falomo96}. From the old
IRAS data there is no evidence for a thermal component in the
far--infrared \citep{impey88}.
\par GC 0109+224 was detected in the past at X-ray bands by satellites
\textit{Einstein}, EXOSAT, ROSAT, and the source is a member of
the RGB catalog \citep{laurent99}, a list of intermediate blazars
with properties smoothly distributed in a large range between the
LBL and HBL subclasses. In the diagnostic diagram
$\alpha_{ro}$-$\alpha_{oX}$ \citep{padovani95}, GC 0109+224
appears close \citep{dennett00} to a prototype of intermediate
blazars like ON 231 \citep[W Com,
][]{tosti02,boettcher02,tagliaferri00}. The HBL and LBL subclasses
exhibit systematically distinct properties (i.e. $F_{X}/F_{rad}$
flux densities ratio, degree of radio core dominance, optical
polarization degree and variability, etc.). The so called
``intermediate'' blazars are critical to clarify the relationship
between these subclasses, and to validate the unified schemes
based on bolometric power. Intermediate blazars often peak in
optical bands (thus are selected and monitored mainly by optical
telescopes), and during some variability phases the break between
the two overall spectral components occurs in the X-ray bands. In
some cases their spectral energy distributions (SEDs) can be still
described well with synchrotron-self-Compton (SSC) models (like
for the HBLs), and this implies, by simple scaling arguments, a
possible inverse Compton (IC) emission peaked in GeV gamma-ray
bands \citep[see, e.g. ][]{stecker96}. In the case of GC 0109+224
EGRET did not detected any gamma-ray emission \citep[phase 1,
][]{fichtel94}, with a rather low upper limit of 0.01 nJy above
100 MeV.
\par GC 0109+224 is regularly monitored by the University of Michigan
Radio Astronomy Observatory (UMRAO), USA \citep[see, e.g.
][]{aller96} and by the Mets\"ahovi Radio Observatory, Finland
\citep[see, e.g. ][]{terasranta98}. In order to search for time
correlations between the optical and radio fluctuations at the
various bands, in Section \ref{sec:radioopt} the updated radio
flux data of UMRAO and Mets\"ahovi are compared with the available
optical data \citep[reported in ][]{ciprini03a}. In Section
\ref{sec:radiovariab} the temporal behavior of the radio light
curves is investigated through well known methods suitable for
unevenly data sets. In Section \ref{sec:sed} we show the
reconstruction of the SED, with the available multiwavelength
data, and a first estimation of the bolometric energy and physical
parameters using a pure SSC model.
%
%
\begin{figure}
\begin{center}
\begin{tabular}{l}\hskip -0.3cm
{\resizebox{\hsize}{!}{\includegraphics{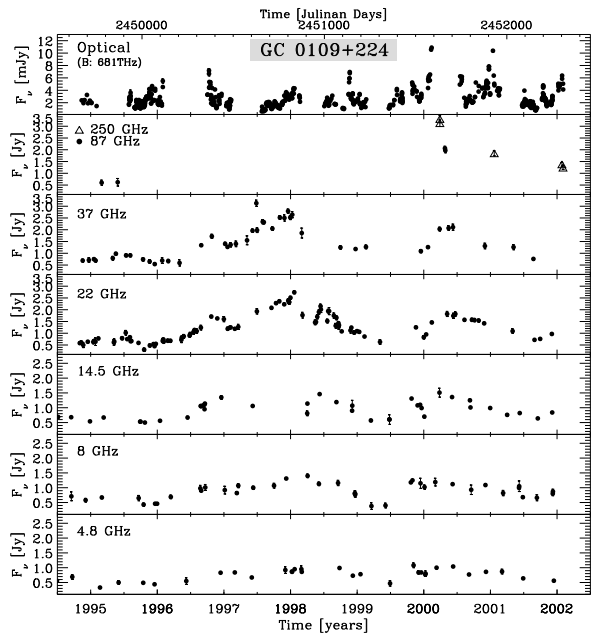}}}\\
\end{tabular}
\end{center}
\vskip -0.1 true cm \caption{The post-1994 portion of the flux
curves in Fig. \ref{fig:clucetutte}, that match the optically best
sampled period. A few observations at mm wavelength (87 and 250
GHz) are also added in this plot. In this detail radio flares
clearly appear to work on considerably longer timescales, with
respect to the optical flares. The relevant emission bump between
the end of 1995 and mid--1999 (clearly visible in the 22 GHz
curve), seems to be the result either of superposition of four
flares departing from base level, or superposition of four peaks
departing from a slower base level bump. Noteworthy is also the
February-April 2000 outburst, because it seems detected (during
the last decreasing phase) at all seven mm--radio frequencies
reported in this plot.} \label{fig:clucebestperiod}
\end{figure}
%
%
\section{Radio-optical correlations}\label{sec:radioopt}
%
The reconstructed optical flux history of GC 0109+224, and the
last seven years of data collected mainly during the Perugia
monitoring program, are compared with the updated radio flux
observations, belong to the Mets\"ahovi (37 GHz and 22 GHz data),
and UMRAO (14.5 Ghz, 8 Ghz and 4.8 GHz data) radio observatories.
At centimetre and millimetre bands (up to about 90 GHz) a monthly
sampling for blazars is in most cases enough. The source was
monitored at an appreciable rate especially at 22 GHz, and in a
significant long-term period at 4.8, 8 and 14.5 GHz. Some
essential properties of the centimetre emission of this blazar
measured at UMRAO are outlined in \citep{aller99}. Mets\"ahovi
data up to 1998 were compared with few literature optical
observations (only 1996-1998), finding a lag of about 400 days
between optical and 22 GHz variations \citep{hanski02}.
%
\begin{figure}
\begin{tabular}{l}\hskip -0.3cm
{\resizebox{\hsize}{!}{\includegraphics{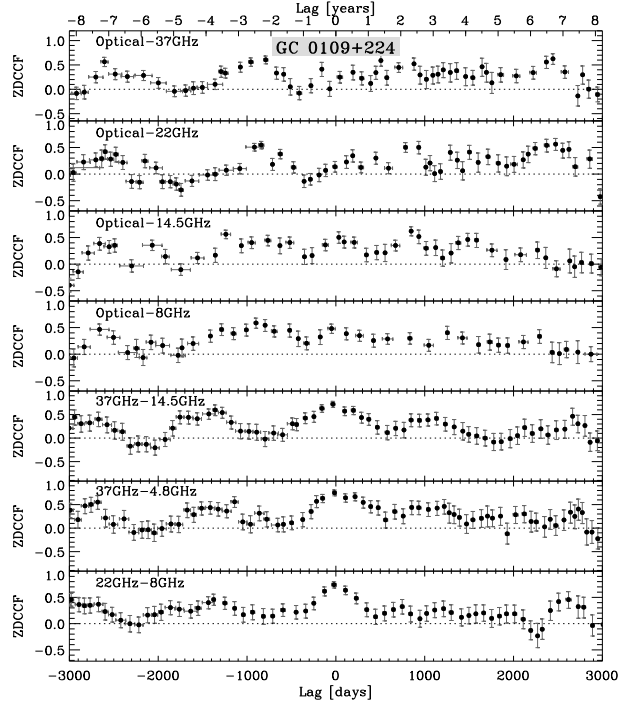}}}\\
\end{tabular}
\vskip -0.3cm \caption{Cross correlation calculated with the
z-transformed discrete correlation function \citep[ZDCF,
][]{alexander97}. Due to the different timescales for the duty
cycles in the radio and optical, and to gaps empty of data in the
optical, no strong evidence of a radio--optical correlation is
found. Weak peaks in the cross--correlations (ZDCCF coefficient
$\simeq 0.5-0.6$) are found around a lag of 900 days (also
positive in the optical--22GHz correlations, see text). On the
other hand, the fluxes at the various radio frequencies appear
well correlated, as expected, with a peak around the zero lag.}
\label{fig:ZDCF}
\end{figure}
%
\par The complete optical (Johnson B band) and radio light curves
of GC 0109+224, are plotted in Fig. \ref{fig:clucetutte}.
Increased activity was observed after 1992 in all five radio
bands. Available optical data are too few prior to 1995, to check
the beginning of enhanced emission at this band. Since 1995
Perugia observations have increased the sampling density,
recording also the larger (and faster) flares, showing well
increased activity, rapid variations, and the same flares
superposition which seems to characterise the radio flux. The flux
history of the source, with millimetre data added, is plotted in
Fig. \ref{fig:clucebestperiod} in more detail. Radio outbursts are
clearly longer then the optical one, and show smaller amplitudes
at lower frequencies. The peak frequency, in which the highest
flux is observed during the flare, seems, at least for the
outburst of March-April 2000 (2000.4, Fig.
\ref{fig:clucebestperiod}), to be around millimetre wavelengths
\citep[as an intense flux of $3.13 \pm 0.04$ and $3.31 \pm 0.06$
Jy at 250 GHz, was detected by the Mambo bolometer at IRAM on 29
March 2000, ][]{bertoldi02}.
%
\begin{figure}
\begin{tabular}{l}\hskip -0.3cm
{\resizebox{\hsize}{!}{\includegraphics{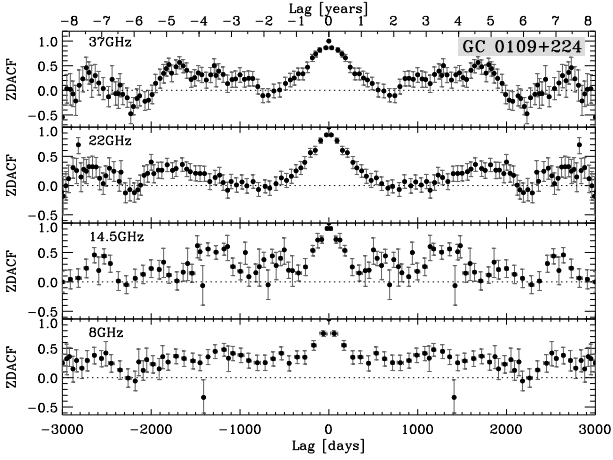}}}\\
\end{tabular}
\vskip -0.3cm \caption{The auto--correlations of the radio flux
curves, calculated with the ZDCF. These do not show any relevant
feature, except for broad and weak ($ZDACF\simeq0.6$) peaks around
the timescale of 3.14-4.06 years (at 14.5 GHz), and around
3.53-4.60 years (at 37 GHz). From 37 GHz to 14.5 GHz, the ZDACF
profile resemble to a shot noise, with a single bump in the
correlation curve within the 2000 days timescale (semi--amplitude
$\simeq$ 2.7 years).} \label{fig:ZDAutoCF}
\end{figure}
%
%
\par The 22 and 37 GHz curves of Fig. \ref{fig:clucebestperiod},
show the faster features in the radio, and five main peaks: around
1996.7 (visible also at 14.5, 8 and 4.8 GHz), around 1997.5,
another around 1998.0 (seen also at 8 GHz), still one before
1998.5 (seen also at 14.5 GHz, while the 37 GHz receiver was
down), and another around 2000.4 (this is seen in all the bands).
Other peaks are easily identifiable (Fig. \ref{fig:clucetutte})
around 1993.0 (from 8 GHz to 37 GHz), and at the end of 1989 (from
4.8 GHz to 14.5 GHz). The more violent optical activity in these
years (Fig. \ref{fig:clucebestperiod}) is characterised by six
main flares of much shorter duration (on the order of weeks-one
month), with faster rise and decline.
\par The noticeable temporal structure in the radio flux curves, between
the end of 1995 and mid--1999, looks like an underling emission
bump spreading over 3.5 years, with four superposed main peaks, or
a superposition of four flares departing from the base level. It's
difficult, even visually, to link this radio bump to a specific
optical peak. The optical emission in this period, apart of an
outburst in Oct. 1996, decreases to a ground level, with a few
flickering low peaks until mid--1997, then rises almost
monotonically with a flickering behaviour, up to beginning of
1998, and then displays another relevant flare only during
November 1998 \citep{ciprini03a}. Another prominent radio outburst
occurred in February-April 2000, and seems detected at all the
seven mm-radio frequencies (from 4.8 to 250 GHz, Fig.
\ref{fig:clucebestperiod}), during the final slow decreasing
phase.
\par Blazars usually exhibit shorter variability timescales, and more
numerous flares in the optical, than in mm-radio bands, and the
optical data are often heavily affected by empty gaps in the
observations (seasonal interruptions, and weather conditions).
\begin{figure}
\begin{tabular}{c}
{\resizebox{8cm}{!}{\includegraphics{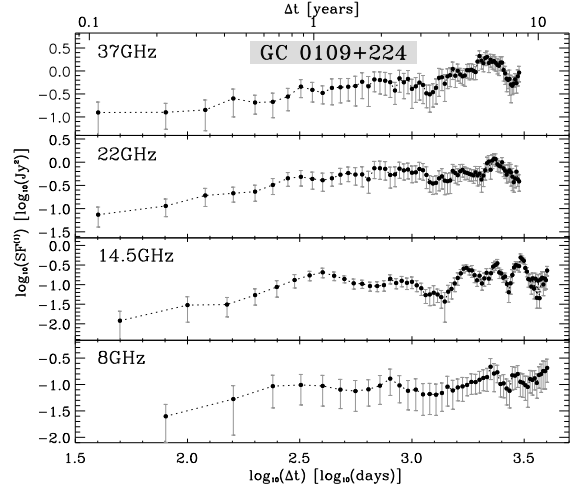}}}
\end{tabular}
\caption{The first order structure functions (SF) of the radio
density flux light curves, in logarithmic scale. For the 37 GHz
flux curve, the mean slope index of the steep part of the SF is
$\beta=0.48\pm 0.08$, and for the 22 GHz curve $\beta=0.65\pm
0.05$ (bin size used is 40 days for both). For the 14.5 GHz series
the SF mean slope is $\beta=0.39\pm 0.14$, and for the 8 GHz data
$\beta=0.52\pm 0.13$, (50 days the bin size used). The time scales
corresponding to deepest drops in the SF are of 3.4-3.8 and
7.4-7.6 years. The shorter lags at which there are hints of a
flattening in the curves, are around 1.1-1.8 years.}
\label{fig:sf}\vspace{-5mm}
\end{figure}
%
This is a problem in the search of correlations. The level of
correlation can be investigated with the Discrete Correlation
Function \citep[DCF,][]{edelson88,hufnagel92}, and the
Interpolated Correlation Function \citep[ICF, ][]{gaskel87},
suitable for discrete unevenly sampled data sets. In the DCF the
number of point per time bin can vary greatly, in the ICF instead,
the interpolation may be unreliable if the curves are
under-sampled. We applied a method that is more robust especially
when the data are few, ensuring that there is a statistically
meaningful number of points in each bin: the Fisher z-transformed
DCF, \citep[ZDCF,][]{alexander97}. This method build data bins by
equal population rather than equal width, and uses Montecarlo
estimations for peaks and uncertainties.
\par For a lag between 789 and 879 days (2.1-2.4
years), no peaks with large values were found in the
optical--radio cross correlation curves (ZDCCF values from 0.50 to
0.62, Fig. \ref{fig:ZDCF}). Other shorter radio-optical delays for
which ZDCCF peaks are recognizable, are around the delay of 190
days (observed in both the correlations with the 37 and 22 GHz
fluxes, but it is not meaningful because $ZDCCF\simeq 0.35$),
around a lag of 33 days ($ZDCCF\simeq 0.50$) between optical-14.5
GHz curves, and around a lag of about 500 days ($ZDCCF\simeq
0.59$) between optical-37 GHz fluxes. Interesting to note is that
the cross--correlations, with a peak around the 2.1-2.4 years lag
range (both positives, delays, and negatives, leads), are found in
all the optical--radio (37, 22, 14.5 and 8 GHz) correlation
curves, (Fig. \ref{fig:ZDCF}). This also appears when applying the
standard DCF method. Hints of optical--radio correlation around
this lag timescale, is also present in the plots of
\citep{hanski02}, while the 350 and 400 days lags cited in this
work (based only on two years of optical data), was not confirmed
by our data.
\par The optical--radio delays found around 2.1-2.4 years,
do not represent strong evidence for a real physical correlation,
because the timescales suggested are very long. Even though the
coefficients are not low (0.5-0.6) and the correlations are found
for all radio frequencies correlations. On the other hand shorter
delays around 33, 190, 500 days are not significant, because they
are found with low value correlations coefficients. To investigate
better the radio--optical delays, it is necessary to increase data
sampling in future optical monitoring programs.
%
\begin{figure}
\begin{tabular}{c}\hskip -0.3cm
{\resizebox{8cm}{!}{\includegraphics{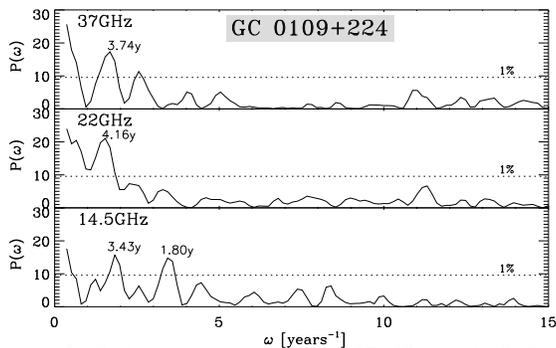}}}\\
\end{tabular}
\vskip -0.5 true cm \caption{Periodogram plots of the 37, 22 and
14.5 GHz flux curves. Dotted lines shows the 1$\%$ false alarm
significance level, under the hypothesis of fluctuations dominated
by Poisson statistics. The timescales corresponding to the peaks
above this level, are 1.80, 3.43, 3.74 and 4.16 years.}
\label{fig:periodogram}
\end{figure}
%
\section{Radio flux variability}\label{sec:radiovariab}
%
The radio flux auto--correlation curves (ZDACF and DACF) of GC
0109+224 in various bands (Fig. \ref{fig:ZDAutoCF}), do not show
very relevant feature, except for broad and weak
($ZDACF\simeq0.6$) peaks around the scale of 3.14-4.06 years in
the 14.5 GHz curve, and around 3.53-4.60 years in the 37 GHz
curve. From 37 GHz to 14.5 GHz the auto--correlation profile
resembles shot noise, with a single bump in the correlation curve
of semi--amplitude $\simeq$ 2.7 years.
\par In Fig. \ref{fig:sf} we plotted the first order structure
functions (SFs) \citep[e.g.][]{simonetti85,hughes92} of the radio
light curves. The SF works in the time domain instead of
frequencies $f$. It measures of the mean difference in the data
train as a function of the separation time $\Delta t$ in the
sampling. In log--log plots the SF of an ideal time series plus
measurement noise, increases and shows an intermediate steep curve
($SF(\Delta t) \propto (\Delta t)^{\beta}$). The slope $\beta$
depends on the nature of the intrinsic variability, and give the
Fourier power law index of the spectrum, (say $P(f)\propto
1/f^{\beta+1}$, where f is the frequency). The values of $\beta $
for GC 0109+224 (Fig\ref{fig:sf}), go from $0.39\pm0.16$ to
$0.65\pm0.08$, meaning an intermediate variability mode, between
the pure flickering (pink noise, $\beta=0$) and the shot noise
(red/brown noise, $\beta=1$). Deep and steep drops in the SF plots
means a little variance, and then a possible signature of
characteristic time scales: it is apparent at about 3.4-3.8 years
and 7.4-7.6 years. Other characteristic timescales are given by
the turnover lag where the SF profile begin to flatten, and in the
plots of Fig. \ref{fig:sf} these are located around 1.1-1.8 years,
even if in our data the flattening is recognizable only with
difficulties.
\par In Fig. \ref{fig:periodogram} are reported the Lomb--Scargle
periodograms \citep{scargle82,horne86} of the radio curves which
present some features. This technique \citep[calculated with a
fast algorithm][]{press89}, analogous to the Fourier analysis for
unevenly sampled data sets, useful to detect possible periodicity,
and typical timescales. The components above the $1\%$ false alarm
threshold, annotated in Fig. \ref{fig:periodogram} (from 1.8 to
4.16 years), are not very significant, and not easily visually
identifiable in the flux light curves.
\par The length of the data record, required to demonstrate a real
recurrent timescale depends on signal--to--noise ratio, systematic
errors, regularity in sampling, nature of the measurements, and
the nature of the underlying variation. Even if the strength of
the autocorrelations found is not high, from the beginning of
1990s our radio sampling is sufficient at all the five bands
observed to consider this discovered hints of timescales, a
starting point in the search for recurrent times with future
observstions. As mentioned above, hints of a sort of periodicity
(or at least of a typical timescale) around 3-4 years, are found
by applying all three methods. Future radio flux data, with
increased sampling and extended observing period, and VLBI
observations may or may not confirm this timescale.
%
\begin{table}
\caption[]{Some roperties and fluxes of GC 0109+224. (1) optical
counterpart \textit{Hipparcos} coordinates \citep{zacharias99};
(2) redshift lower limit (Falomo 1996); (3) 82 cm flux density
\citep{douglas96}; (4) 20 cm flux density \citep{owen80}; (5) mean
flux densities from this work; (6) 87 GHz flux from Mets\"ahovi;
(7) 1.2 mm Mambo-IRAM detection \citep{bertoldi02}, (8) IRAS
far-IR flux at 60 $\mu$m \citep{impey88}; (9) mid-IR flux in L
band \citep{odell78}; (10) mean optical flux in R Cousins band
\citep{ciprini03a}; (11) mean optical flux in B Johnson band from
historical light curve \citep{ciprini03a}; (12) optical (B band)
min and max degree of liner polarization
\citep{valtaoja93,takalo91}; (13) X-ray flux at 2 keV by Einstein
Observatory (HEAO-2) IPC and MPC instruments
\citep{owen81,ledden85,maraschi86}; (14) X-ray flux at 1 keV by
CMA instrument on EXOSAT \citep{maraschi88,giommi90,reynolds99};
(15) X-ray flux at 1 keV by PSPC instrument of ROSAT
\citep{neumann94,brinkman95,reich00} and spectral index
\citep{kock96,laurent99}; (16) flux limit above 100 MeV by EGRET
(see Tab. \ref{tab:propgamma}), \citep{fichtel94}.}
\label{tab:propgeneral} \vspace{-0.2cm}
\centering {}
\par
\scriptsize{
\begin{tabular}{lccc}
\vspace{-3mm} \\
\hline \hline
\vspace{-3mm} \\
Quantity & Value & Epoch & Ref.   \\
\hline \hline
R.A.(J2000.0)$_{opt}$         & 01h 12m 05.8238s         & ... &(1)  \\
Dec.(J2000.0)$_{opt}$  & +22$^\circ$ 44' 38.798'' & ... &(1) \\
z                      &  $>0.4$                  & ... &(2)\\
F$_{\nu}$(0.365 GHz)    &$0.279\pm0.021$ Jy& ...         &(3)\\
F$_{\nu}$(1.484 GHz)    &$0.36\pm0.2$ Jy   & Jan78       &(4)\\
$<$F$_{\nu}$(4.8 GHz)$>$& 0.56 Jy          & Feb80-Dec01 &(5)\\
$<$F$_{\nu}$(8 GHz)$>$  & 0.67 Jy          & May78-Dec01 &(5)\\
$<$F$_{\nu}$(14.5 GHz)$>$ & 0.66 Jy        & Sep79-Dec01 &(5)\\
$<$F$_{\nu}$(22 GHz)$>$ & 1.07 Jy          & Jan85-Dec01 &(5)\\
$<$F$_{\nu}$(37 GHz)$>$ & 1.05 Jy          & Jul84-Aug01 &(5)\\
F$_{\nu}$(87 GHz)       & $2.06\pm0.09$ Jy & 2000 Apr 26 &(6)\\
F$_{\nu}$(250 GHz)      & $3.31\pm0.06$ Jy & 2000 Mar 29 &(7)\\
F$_{\nu}$(60 $\mu$m)    & 188 mJy          & 1983        &(8)\\
F$_{\nu}$(3.5 $\mu$m)   & 11.8 mJy         & Jan77       &(9)\\
$<$F$_{\nu}$(0.64 $\mu$m)$>$& $3.8\pm0.2$  & Nov94-Feb02 &(10)\\
$<$F$_{\nu}$(0.44 $\mu$m)$>$& $2.4\pm0.3$  & Nov76-Feb02 &(11)\\
P($B$)$_{min/max}$      & 0.7\% / 29.69 \% &   ...       &(12)\\
F$_{\nu}$(2 keV)        &  $0.15$ $\mu$Jy  & Jul 1979    &(13)\\
F$_{\nu}$(1 keV)        &  $0.21$ $\mu$Jy  & Aug 1984    &(14)\\
F$_{\nu}$(1 keV)        &  $0.26\pm 0.08$ $\mu$Jy & Aug 1990 &(15)\\
$\alpha_{\textrm{x}}$   &  $1.96\pm 0.25$         & Aug 1990 &(15)\\
F$_{\nu}$($>100$ MeV)   & $< 0.01$ nJy     & 1992        &(16)\\
\hline\hline
\end{tabular}
}
\end{table}
\normalsize
%
%
\begin{table*}
\caption[]{EGRET viewing periods (v.p.) and flux upper limits
(u.l.) of GC 0109+224. Related quasi-simultaneous radio and
optical fluxes from this work, are also reported in the Table when
available. References: (a) EGRET integral flux density upper limit
above 100 MeV (first EGRET catalog, phase 1) for three v.p.
\citep{fichtel94}; (b) the only simultaneous optical data
available \citep[][]{valtaoja93}, points out a low brightness and
polarization degree $5.20\pm0.61$; (c) u.l. given by the summed
exposures of phase 1 pointings, and the other EGRET v.p. of the GC
0109+224 region on phase 3 and cycle 4 (1994-95)
\citep{hartman99}. Any new pointing of the region was done during
the EGRET phase 2 \citep[Nov.1992-Sept.1993, ][]{thompson95}.
During EGRET cycle 8, at least one more pointing was available
(2000, Apr.18-Apr.25) but our source was on the edge of the field
of view. Throughout all the EGRET v.p. the source had a relatively
low (sub-Jy) radio emission.} 
{\resizebox{16cm}{!}{\includegraphics{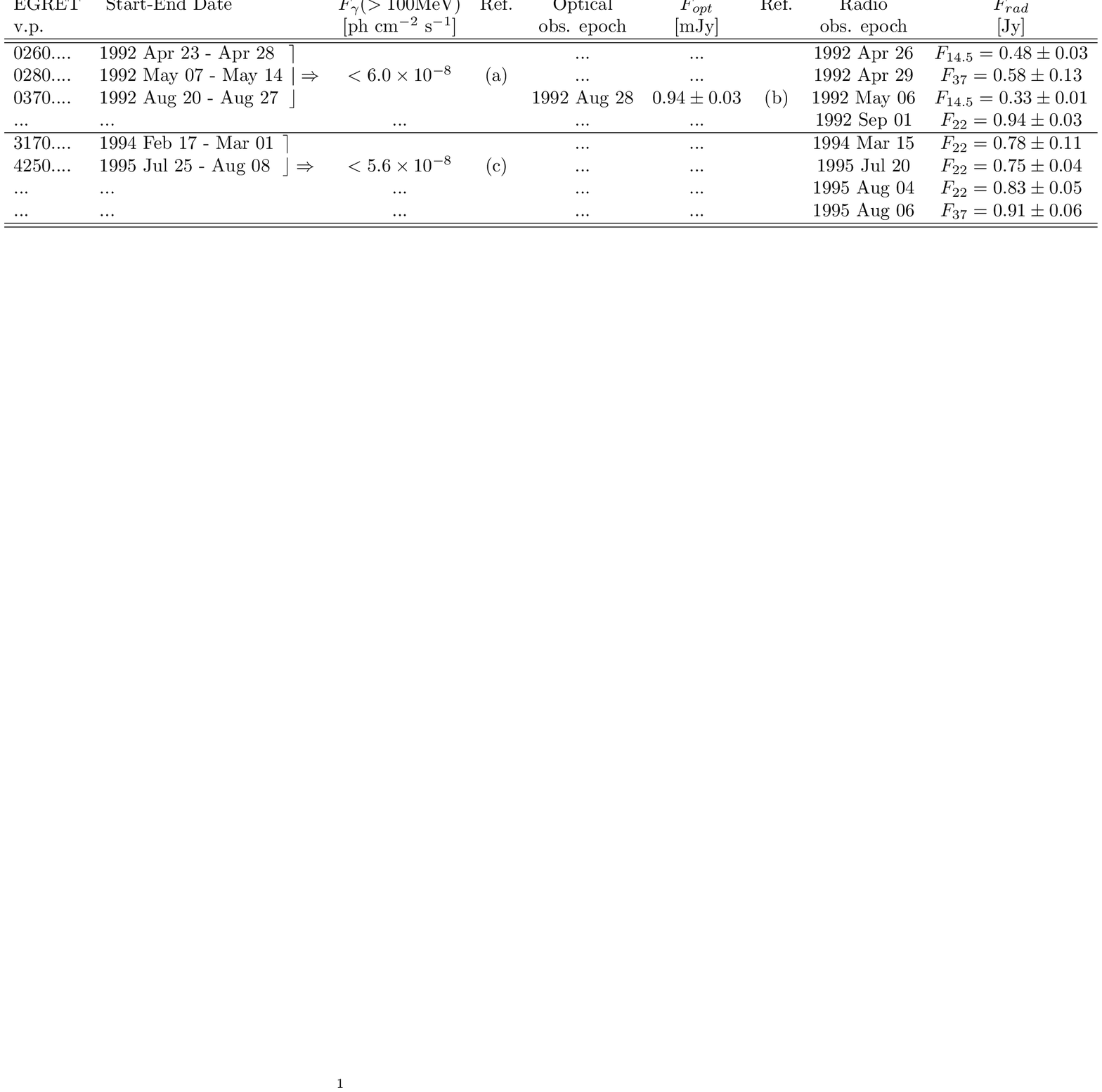}}}
\label{tab:propgamma}\vspace{-5mm}
\end{table*}
%
\section{The Spectral Energy Distribution}\label{sec:sed}
\begin{figure*}
\begin{center}
\begin{tabular}{l}
\hspace{-6mm}
{\resizebox{\hsize}{!}{\includegraphics{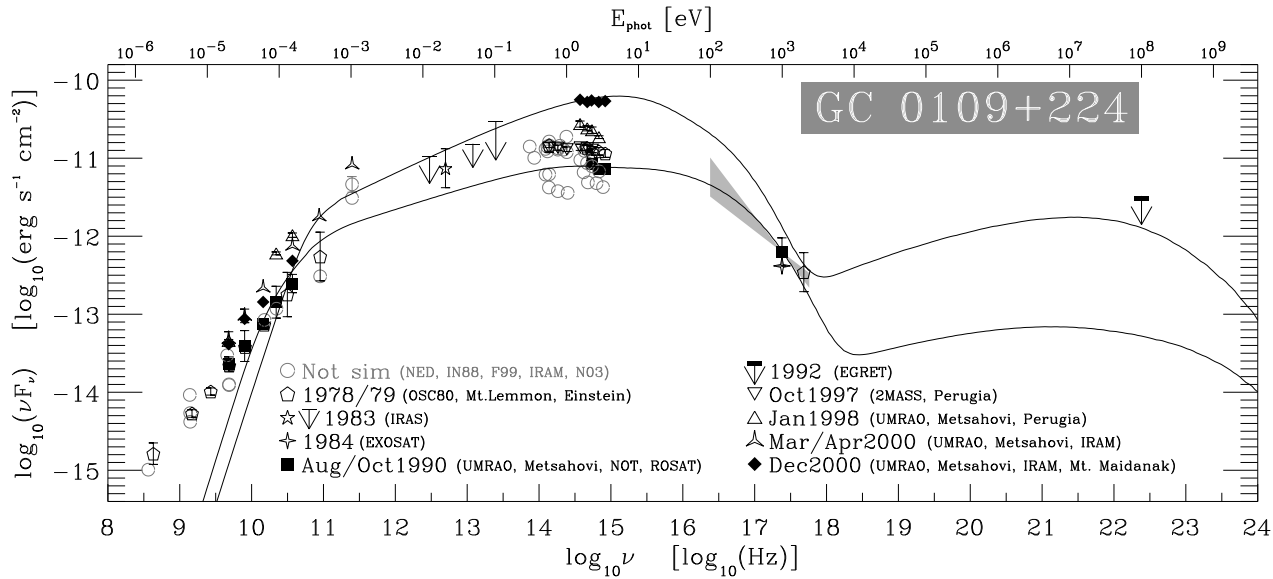}}}\\
\end{tabular}
\end{center}
\vskip -0.5 true cm \caption{The SED of GC 0109+224 assembled with
the available multiwavelength data (open and filled symbols),
upper limits (arrows), and ROSAT slope (bow-tie), at different
epochs. Two SSC model fit attempts (black lines) are used, for a
low emission state and an high state. The SED is consistent with a
pure smooth synchrotron continuum, peaking between the near-IR,
optical, near-UV, range. EGRET upper limit above 100 MeV is rather
low. There is no evidence of gamma-ray emission bump, and the few
X-ray detections suggests a downward trend. Data are grouped for
no-simultaneous epochs with the same symbol. Instruments or
references are indicate in parentheses. Data from this work
(UMRAO, Mets\"{a}hovi and Perugia observatories), NED database,
from \citep[][ OSC80]{owen80}, Mt. Lemmon \citep{puschell80}, from
\citep[][ F99]{fan99} and \citep[][ N03]{nesci03}, Einstein IPC
\citep{owen81,ledden85}, IRAS \citep[][ IN88]{impey88}, EXOSAT CMA
\citep{maraschi88,giommi90,reynolds99}, NOT \citep{valtaoja93},
ROSAT PSPC \citep{neumann94,brinkman95,kock96,reich00}, EGRET
\citep{fichtel94,hartman99}, 2MASS \citep{skrutskie97}, IRAM Mambo
\citep[Mar00, Jan01 and Jan02, ][]{bertoldi02} and Mt. Maidanak
\citep{ciprini03a}. The two lines referring to the SSC fit
attempts for the flaring state of December 2000 and the low state
around the ROSAT detection on 1990, are performed with the
following parameters. \textbf{August/October 1990 low state}:
$\gamma_{min}=60$, $\gamma_{max}=1.7 \times 10^{6}$,
$\gamma_{break}=5\times 10^{3}$, $p_{1}=2.4$, $p_{2}=2.5$,
$k=0.015$, $B=0.45$Gauss, $R=9.5 \times 10^{16}$cm,
$\mathcal{D}=13$. \textbf{December 2000 high state}:
$\gamma_{min}=20$, $\gamma_{max}=5\times 10^{5}$,
$\gamma_{break}=7\times 10^{3}$, $p_{1}=2.1$, $p_{2}=2.2$,
$k=0.015$, $B=1.6$Gauss, $R=8.5 \times 10^{16}$cm,
$\mathcal{D}=15$. Redshift $z=0.4$ is assumed \citep{falomo96}.}
\label{fig:sed0109}
\end{figure*}

GC 0109+224 is a bright X-ray point source ($\nu
F_{\nu}(1\textrm{~keV}) \gtrsim10^{-13}$ erg cm$^{-2}$ s$^{-1}$),
known since the observation of HEAO-1 A2 experiment
\citep{dellaceca90} in Jan. 1978. The spectral energy distribution
(SED) assembled with the available historical multiwavelength
data, is shown in Fig. \ref{fig:sed0109} (references for data in
figure's caption), and is compatible with a smooth continuum given
by a pure synchrotron component, like the classical BL Lac
objects. In Table \ref{tab:propgeneral} we summarize properties
and available multiwavelength fluxes of GC 0109+224. The SED data
profile in Fig. \ref{fig:sed0109}, points out a synchrotron peak
in the frequency range around the near-infred, optical (and
possibly near-UV) range. This strengthens the assumption that GC
0109+224 is an LBL or intermediate blazar, \citep[as suggested for
example in ][]{laurent99,dennett00,bondi01}.
\par Unfortunately, a part of one rather low EGRET upper limit, no
data is available to check the presence of the inverse Compton
(IC) emission bump (the few and old X-ray observations seems to
have a synchrotron origin). The 1 keV flux, measured by ROSAT PSPC
instrument in August 1990, is $F_{X}=0.26\pm 0.08 \mu$Jy, with
$\alpha_{X}=1.96 \pm 0.25$ (Table \ref{tab:propgeneral} and Fig.
\ref{fig:sed0109}). This suggests a decreasing soft-X trend, in
the $\nu F_{\nu}$ representation. The 1 keV flux by EXOSAT CMA
instrument in August 1984, is approximately the same ($F_{X}=0.21
\mu$Jy), while the 2keV flux by \textit{Einstein} Observatory
(HEAO-2) IPC and MPC instruments in July 1979, is
$F_{X}=0.15~\mu$Jy. These detections seem to prove the highest
energy synchrotron tail is emitted in the soft-X ray band (as is
commonly found in intermediate blazars).
\par IRAS measured a flux density of 188 mJy at 60 $\mu$m and upper
limits at the other its bands \citep{impey88}. As cited above GC
0109+224 seems characterized by a very strong (2-3 Jy) millimetre
flux (a part of calibration uncertainties). On March 29, 2000, the
IRAM Mambo bolometer measured a flux of 3.13 and 3.31 Jy at 1.2 mm
(250 GHz), \citep[and all other few available measurements are
above 1 Jy, ][]{bertoldi02}. While on April 26, Mets\"{a}hovi
detected a flux above 2 Jy at 87 GHz (Fig.
\ref{fig:clucebestperiod} and Tab. \ref{tab:propgeneral}). No
simultaneous optical data were available during this period in
which the source is not visible on night, but February 2000
observations suggest a relevant high emission state
(F$_{\nu}$(0.44 $\mu$m)$\simeq 16$ mJy). According to
\citep{valtaoja96,jorstad01}, $\gamma$-ray flares are considered
connected to the rising or high state of the blazar millimetre
emission. GC 0109+224 seems on a rising phase during September
1992 as showed by the radio flux curves, but it was not detected
in $\gamma$-rays by EGRET. This could contradict the cited
hypothesis, or the $\gamma$-rays were of short duration, or again
probably the sensitivity was inadequate. The integral flux density
upper limit (u.l.) above 100 MeV is rather low: the limit given in
the first EGRET catalog (phase 1, Apr. 1991 - Nov. 1992) is
$F_{\gamma}<6\times
10^{-8}~\textrm{photons}~\textrm{cm}^{-2}~\textrm{s}^{-1}$,
approximatively equivalent to 1 nJy. The low optical brightness
and the small radio fluxes (sub-Jansky), reported in Table
\ref{tab:propgamma}, in correspondence with the EGRET viewing
periods, point out a low or, in any case, non-flaring states. For
details on the EGRET viewing periods pertaining to GC 0109+224,
and corresponding available optical and radio fluxes, see Table
\ref{tab:propgamma}.
\par We tentatively fitted the few available data with a pure one–-zone homogeneous
leptonic SSC model, in two representatives (low and high) states
of the source (continuous lines in Fig. \ref{fig:sed0109} for the
August-October 1990 quiescent state, and the December 2000 flaring
state). The multiwavelength SED data are grouped in epochs in Fig.
\ref{fig:sed0109} (but we remark that this is only a rough
incongruent and temporally broad regroupment of the data). The
applied model assumes a blob of dimension $R$ embedded in a
tangled isotropic magnetic field of mean intensity $B$, subjected
to a continuous injection of shocked relativistic electrons with a
break power law energy distribution, exponentially damped:
$Q_{inj}(\gamma)=Q_{0}\gamma^{-p}\exp(-\gamma/(k\gamma_{max}))$
[cm$^{-3}$ s$^{-1}$], between the energies (Lorentz factors)
$\gamma_{min}$ and $\gamma_{max}$. This emitting region moves in a
relativistic jet, approximatively aligned with our line of sight
and fed by an accreting supermassive black hole. The mechanism is
described using a one--dimensional and time--dependent kinetic
equation for the electron particle distribution, and the ensemble
synchrotron spectrum is convolved with the calculated
distribution. The IC spectrum results from the interaction of the
distribution with the synchrotron photon field. The produced
spectra were transformed to the frame of the observer (placed at
an angle $\theta$ respect to jet direction), using the
relativistic Doppler beaming bulk factor
$\mathcal{D}=\left((1+z)\Gamma(1-\beta \cos \theta)\right)^{-1}$.
The model is described for example in
\citep{ciprinitosti03c,ciprini03d}. The values of the parameters
adopted for the two SED parameterisation of GC 0109+224, are
reported in the figures caption. The steep injection ($p > 2$) and
the exponential damping, creates a few high energy electrons
(important for the IC emission). Moreover the relatively high
value of $B$ (0.45-1.6 G), means a stronger magnetic field energy
density $ U_{B}=B^{2}/(8\pi ) $, and a smaller radiation to
magnetic energy ratio $U_{rad}/U_{B}=L_{IC}/L_{syn}$, redicing the
self-Compton flux. Indeed in our modelling (Fig.
\ref{fig:sed0109}) the IC component appears depressed. The few
multiwavelegth data available do not contain any hint of an
high-energy spectral component. The values used for the physical
parameters partially take in account radio flux data in the SED.
For example lower values of $\gamma_{min}$ could fit better the
lower frequency radio data, but the IC emission is further
depressed. The synchrotron spectral shape indicated by the
available data, appears also to be suitable for a fit with
logarithmic parabolic functions \citep[see, e.g.
][]{giommi02,sambruna96}. Probably, as stated for the flux curves,
the dynamics and the regions involved in the high energy emission
are different from the radio ones. Anyway the usual neglect of the
radio data points in blazar SED modelling is not ever justified.
%
%
\par SED data and modelling (Fig. \ref{fig:sed0109}), allude to
a synchrotron peak ranging between the near-IR, for the quiescent
states, and the optical (perhaps also near-UV) bands for the
flaring states.
This is consistent with the view of GC 0109+224, as a blazar at
the border of the LBL and intermediate subclasses. As a classical
BL Lac, the SED is represented well with our pure leptonic SSC
model. Therefore, with simple scaling considerations, the IC
emission can be seen as the synchrotron component upshifted by
$\gamma_{max}^{2}$ (with $\gamma_{max} \simeq 10^{5} $ in this
case). The peak in the optical, in SSC descriptions, can predict a
possible IC peak in GeV $\gamma$-ray energies through the
following similarity relation: $(\nu_{GeV}F_{GeV})/L_{IC} \simeq
(\nu_{opt}F_{opt})/L_{syn}$, (as the the synchrotron X-ray end
tail is related to the TeV $\gamma$-ray tail in the HBLs
\citep{stecker96}). $L_{syn}$ and magnetic fields intensity might
be high for this object, reducing the IC dominance. However the
strong millimeter flux, the peak of the SED in near-IR-optical
bands, and the brightness in X-ray frequencies, might suggest the
possibility of $\gamma$-ray emission. This emission may be
detectable by the next generation of $\gamma$-ray satellites with
improved sensitivity (e.g. AGILE and GLAST). Moreover we note the
possible correlation of GC 0109+224 with one of the highest energy
cosmic-rays events ever detected, $E = (1.7-2.6)\times 10^{20}$eV,
on Dec. 3, 1993 \citep{farrar98}. In the scenario evading the
GZK-cutoff, the primaries produced by particle acceleration in a
blazar, should travel extragalactic distances not deflected,
pointing directly to its source, and under this assumption the
probability of a coincidental alignment of that event with GC
0109+224 was only $0.5\%$ \citep{farrar98}.
%
\section{Summary and Conclusions}
In this work we have reported and analysed the largest amount of
radio and optical data ever published on the BL Lac object GC
0109+224, collected over more than twenty years, and we have
reconstructed also the most complete multiwavelength SED
available. Some results are found, but also open questions are
arisen. GC 0109+224 is an example of a blazar which can produce
strong outburst at high radio frequencies (2-3 Jy at $\nu \gtrsim
20$ GHz), but can remain relatively quiet ($\simeq 0.3$Jy GB6
survey) for long periods. The flux density variability is
characterized by intermittent behaviour, with a not regular
alternation of relatively large amplitude flares, and flickering
phases, both in radio and optical regimes. The source varies over
all timescales sampled (days, months, years). Optical flares are
clearly much faster with respect to the radio-mm bands, and some
particularly large optical outbursts do not have obvious
counterparts at mm and radio wavelengths. This could suggest
additional or different emitting components responsible for the
radio and the optical emission, and/or more rapid cooling of the
synchrotron particles at higher frequencies. Moreover radio
flares, due to the longer duty cycles, can blend, inhibiting the
identification of radio counterparts of the optical flares.
\par The radio light curves appear well correlated at the different
frequencies, around the zero lag. The radio--optical
cross--correlation peaks, found in all the bands around the lag of
2.1-2.4 years, are extended and have low values (ZDCCF $< 0.62$).
The complexity, spread and low values of the correlation peaks,
the impossibility to visually recognize the lags in the light
curves, the very different duty-cycles, and especially such long
delays, suggest that there is not a real correlation, and no
physical meaning in the two-year lag. Other shorter delays around
33, 190, and 500 days are not significant because they are found
with very small correlation coefficients. Fig. \ref{fig:ZDCF}
shows also that the signal is weak at zero time lag (between
optical and 37-22 GHz curves). This means that strong events in
the optical bands do not have a simultaneous recognizable
counterpart in the radio bands. Maybe additional processes and/or
different regions are responsible for the emission at different
frequencies (optical photons could be generated into jet
components of different size and dynamics, in correspondence to
the radio emission). A certain level of confusion is undoubtedly
created by induced causes (i.e. the large empty gaps in the
optical light curve data). Only with continuous monitoring over
longer periods, with reduced gaps, will clarify the existence of
physical radio--optical delays in this source.
\par The DCF and SF shapes reflect the underlying nature of the process
that created the radio variability. Both methods demonstrate, for
GC 0109+224, a fluctuation mode between the flickering and the
shot noise (i.e. $P(f)\propto 1/f^{\alpha}$), with $1.39< \alpha
<1.65$. A similar behaviour was found also for the optical
emission \citep{ciprini03a,ciprini03b} of this source. This power
spectrum is characteristic of a random walk. In the case of the
best sampled 22 GHz light curve, the value $\alpha = 1.65$ found,
is in strict agreement with the $5/3$ value of the fully develop
turbulence in the scalar theory of Kolmogorov. Such variability
could be related to fluctuations of the magnetic field, or of the
bulk motion velocity of the emitting regions inside the jet.
\par  There is no evidence from our data, for long
term periodic variations with a fixed period, but typical
timescales in the ranges of $\sim 3.2-4.5$ years are implied, with
all the applied methods (DCF, SF, periodogram). The more rapid
flickering synchrotron peaks, seem superimposed to long term
trends in the radio. This could also be the result of flares
blending with long cooling times, as mentioned above.
\par The increased activity and mean flux level, well observed in all
the radio--optical bands, suggest that the same emission mechanism
is responsible for radiation in both spectral regimes, i.e. the
synchrotron radiation from shocked plasma into the jet, as seen
also in the SED. Synchrotron emission in GC 0109+224 is probably
dominant and powerful ($L_{syn}/L_{IC} > 1 $), as suggested by the
high degree of the polarization, and by the absence of emission
lines ($z$ is still undetermined). The relatively high degree of
linear polarization observed could mean a weaker connection with
the usual depolarization effects, which is a common affliction of
blazars jets (like influence of a luminous host galaxy). The
overall SED of GC 0109+224, shows a smooth synchrotron continuum
peaked in the near-IR-optical range (as showed by intermediate
blazars). Also our homogeneous SSC modelling suggests a relatively
high magnetic field and synchrotron luminosity, diminishing the
self-Compton radiation, even thought the multiwavelength data
reported are insufficient to fully constrain the models. In
particular GC 0109+224, suffers from substantial lack of data in
millimetre sub-mm and infrared bands, to check the importance of a
possible thermal emission component (and also the possibility of
external-Compton contributions).
\par Despite of the common deficiency of infrared data, and difficulties
for far and mid-IR blazar monitoring, we suggest at least
millimetre observations, due to the intense radiation which GC
0109+224 seems produce at these frequencies. X-ray observations of
this blazar are also strongly encouraged, to check the presence of
the high energy component in the spectrum. The high energy
predictions for an intermediate blazar suffer uncertainties that
become relevant in high energy tails, even when only using
leptonic models \citep{boettcher02}. Moreover the X-ray data are
insufficient to permit any prediction about TeV gamma-rays from
this object (due to the suggested $z> 0.4$, TeV emission might be
interesting for studies on the extragalactic background light by
absorption cutoffs). However millimetre brightness, and
synchrotron emission peaked at optical frequencies, could imply a
GeV gamma-ray radiation detectable by the next-generation of
gamma-ray space telescopes. An increased observing effort for this
source, especially at X-ray bands and beyond \citep[for example GC
0109+224 is just scheduled to be observed by Integral][]{pian02}),
together with a better monitoring in radio-mm-optical bands, will
clarify some of the questions that have arisen in our data and
analysis. In this view our data and work will be useful database.
%
%
\section{Acknowledgments}
%
We wish to thank Kyle Augustson for useful comments. The Metsähovi
Radio Observatory is an institute of the Helsinki University of
Technology, partially financed by the Academy of Finland. The
UMRAO is supported by the US National Science Foundation, and by
funds from the University of Michigan. The optical monitoring
program of the Perugia University Observatory was partly supported
by the Italian MIUR under grant Cofin2002. The European Institutes
belonging also to the ENIGMA collaboration acknowledge EC funding
under contract HPRN-CT-2002-00321. This research has made use also
of: SIMBAD database (CDS, Strasbourg), NASA/IPAC NED database (JPL
CalTech and NASA), HEASARC database (LHEA NASA/GSFC and SAO),
NASA's ADS digital library.
%
%
\bibliographystyle{aa}

\begin{thebibliography}{}



\bibitem[{Alexander (1997)}]{alexander97} Alexander, T. 1997, in Astronomical Time Series, Eds. Maoz,
Sternberg \& Leibowitz, Dordrecht: Kluwer, p. 163

\bibitem[{Aller et al. (1999)}]{aller99} Aller, M. F., Aller, H. D., Hughes, P. A., \& Latimer G.
E. 1999, ApJ, 512, 601

\bibitem[{Aller et al. (1996)}]{aller96} Aller, H.~D., Aller, M.~F., \& Hughes, P.~A.\ 1996, in ASP Conf.~Ser.~110: Blazar
Continuum Variability, p. 208

\bibitem[{Bertoldi (2002)}]{bertoldi02} Bertoldi, F., 2002, private communication

\bibitem[{Brinkmann et al. (1995)}]{brinkman95} Brinkmann, W., Siebert, J., Reich, W., et al. 1995, A\&AS, 109, 147

\bibitem[{B{\" o}ttcher et al. (2002)}]{boettcher02} B{\" o}ttcher, M., Mukherjee, R., \& Reimer, A.\ 2002, ApJ, 581,
143

\bibitem[{Bondi et al. (2001)}]{bondi01} Bondi, M., March\~{a}, M. J. M., Dallacasa, D., \& Stanghellini, C.
2001, MNRAS, 325, 1109

\bibitem[{Ciprini et al. (2003a)}]{ciprini03a} Ciprini, S., Tosti, G.,
Raiteri et al. 2003a, A\&A, 400, 487

\bibitem[{Ciprini et al. (2003b)}]{ciprini03b} Ciprini, S., Fiorucci, M.,
Tosti G., \& Marchili, N. 2003b, ASP Conf. Series, 299, 265

\bibitem[{Ciprini \& Tosti (2003c)}]{ciprinitosti03c} Ciprini S., \& Tosti G. 2003c,
ASP Conf. Series, 299, 269

\bibitem[{Ciprini (2003d)}]{ciprini03d} Ciprini, S. 2003d, New
Astronomy Rev., 47, 709

\bibitem[{Della Ceca et al. (1990)}]{dellaceca90} Della Ceca, R., Palumbo, G. G. C., Persic, M., et al. 1990, ApJS, 72, 471

\bibitem[{Dennett-Thorpe et al. (2000)}]{dennett00} Dennett-Thorpe, J., \& March\~{a}, M. J. 2000, A\&A, 361, 480

\bibitem[{Douglas et al. (1996)}]{douglas96} Douglas, J.~N., Bash,
F.~N., Bozyan, F.~A., Torrence, G.~W., \& Wolfe, C.\ 1996, AJ,
111, 1945

\bibitem[{Edelson \& Krolik (1988)}]{edelson88} Edelson, R. A., \& Krolik, J. H. 1988, ApJ, 333, 646

\bibitem[{Falomo (1996)}]{falomo96} Falomo, R. 1996, MNRAS, 283, 241

\bibitem[{Fan (1999)}]{fan99} Fan, J. H. 1999, astro-ph/9910269

\bibitem[{Farrar \& Biermann (1998)}]{farrar98} Farrar, G. R., \& Biermann, P. L. 1998,
Phys. Rev. Lett., 81, 3579

\bibitem[{Fey \& Charlot (2000)}]{fey00} Fey, A. L., \& Charlot P. 2000, ApJS, 128, 17

\bibitem[{Fichtel et al. (1994)}]{fichtel94} Fichtel, C. E., Bertsch, D. L., Chiang, J., et al. 1994, ApJS, 94, 551

\bibitem[{Gaskell \& Peterson (1987)}]{gaskel87} Gaskell, C.~M.~\&
Peterson, B.~M.\ 1987, ApJS, 65, 1

\bibitem[{Giommi et al. (2002)}]{giommi02} Giommi, P. et al. 2002,
in Blazar Astroph. with BeppoSAX \& Other Obs., ASI spec. publ.,
Rome, p. 63

\bibitem[{Giommi et al. (1990)}]{giommi90} Giommi, P., Barr, P., Garilli, B., Maccagni, D., \& Pollock, A. M. T., 1990, ApJ, 356, 432

\bibitem[{Hanski et al. (2002)}]{hanski02} Hanski, M.\~T., Takalo, L.\~O., \& Valtaoja, E.\ 2002, A\&A, 394, 17

\bibitem[{Hartman et al. (1999)}]{hartman99} Hartman, R.~C., Bertsch, D. L., Bloom, S.
D., et al. 1999, ApJS, 123, 79

\bibitem[{Horne \& Baliunas (1986)}]{horne86} Horne, J. H., \& Baliunas, S. L. 1986, ApJ, 302, 757

\bibitem[{Hufnagel \& Bregman (1992)}]{hufnagel92} Hufnagel, B. R., \& Bregman, J. N. 1992, ApJ, 386, 473

\bibitem[{Hughes et al. (1992)}]{hughes92} Hughes, P. A., Aller, H. D., \& Aller, M. F. 1992, ApJ, 396, 469

\bibitem[{Impey \& Neugebauer (1988)}]{impey88} Impey, C. D. \& Neugebauer, G. 1988,
AJ, 95, 307

\bibitem[{Jorstad et al. (2001)}]{jorstad01} Jorstad, S.~G., Marscher, A.~P., Mattox, J.~R., Aller, M.~F., Aller, H.~D.,
Wehrle, A.~E., \& Bloom, S.~D.\ 2001, ApJ, 556, 738

\bibitem[{Kock et al. (1996)}]{kock96} Kock, A., Meisenheimer, K., Brinkmann, W., Neumann, M., \& Siebert, J. 1996, A\&A, 307, 745

\bibitem[{Laurent-Muehleisen et al. (1999)}]{laurent99} Laurent-Muehleisen, S. A., Kollgaard, R. I., Feigelson, E. D., Brinkmann, W., \& Siebert, J. 1999, ApJ, 525, 127

\bibitem[{Ledden \& Odell (1985)}]{ledden85} Ledden, J. E., \& Odell, S. L. 1985, ApJ, 298, 630

\bibitem[{Maraschi \& Maccagni (1988)}]{maraschi88} Maraschi, L., \& Maccagni, D., 1988, MemSAIt, 59, 277

\bibitem[{Maraschi et al. (1986)}]{maraschi86} Maraschi, L., Ghisellini, G., Tanzi, E.~G., \& Treves, A.\ 1986,
ApJ, 310, 325

\bibitem[{March\~{a} et al. (1996)}]{marcha96} March\~{a}, M. J. M., Browne, I. W. A., Impey, C. D., \& Smith, P. S. 1996, MNRAS, 281, 425

\bibitem[{Mead et al. (1990)}]{mead90} Mead, A. R. G., Ballard, K. R., Brand, P. W. J. L., Hough, J.
H., Brindle, C., \& Bailey, J. A. 1990, A\&AS, 83, 183

\bibitem[{Nesci et al. (2003)}]{nesci03} Nesci, R., Sclavi, S.,
Maesano, et al. 2003, MemSAIt, 74, 169

\bibitem[{Neumann et al. (1994)}]{neumann94} Neumann, M., Reich, W., F\"{u}rst, E., et al. 1994, A\&AS, 106, 303

\bibitem[{Odell et al. (1978)}]{odell78} Odell, S. L.; Puschell, J. J.; Stein, W. A., et al 1978, ApJ, 224, 22

\bibitem[{Owen \& Mufson (1977)}]{owen77} Owen, F. N., \& Mufson, S. L. 1977, AJ, 82, 776

\bibitem[{Owen et al. (1981)}]{owen81} Owen, F. N., Helfand, D. J., \& Spangler, S. R. 1981, ApJ, 250, 550

\bibitem[{Owen et al. (1980)}]{owen80} Owen, F. N., Spanger, S. R., \& Cotton, W. D. 1980, AJ, 85, 351

\bibitem[{Padovani \& Giommi (1995)}]{padovani95} Padovani, P.~\& Giommi, P. 1995, ApJ, 444, 567

\bibitem[{Pauliny-Toth et al. (1972)}]{pauliny72} Pauliny-Toth, I. I. K., Kellermann, K. I., Davis, M. M., Fomalont, E. B., \& Shaffer, D. B. 1972, AJ, 77, 265

\bibitem[{Pian (2002)}]{pian02} Pian, E. 2002, Blazar TOO, Integral Ann. of Opp. (AO-1)

\bibitem[{Press \& Rybicki (1989)}]{press89} Press, W. H., \& Rybicki, G. B. 1989, ApJ, 338, 277

\bibitem[{Puschell \& Stein (1980)}]{puschell80} Puschell, J. J., \& Stein, W. A. 1980, ApJ, 237, 331

\bibitem[{Reich et al. (2000)}]{reich00} Reich, W., Fuerst, E., Reich, P., et al. 2000, A\&A, 363, 141

\bibitem[{Reynolds et al. (1999)}]{reynolds99} Reynolds, A. P., Parmar, A. N., Hakala, P. J., et al. 1999, A\&AS, 134, 287

\bibitem[{Sambruna et al. (1996)}]{sambruna96} Sambruna, R.~M., Maraschi, L., \& Urry, C.~M.\ 1996, ApJ, 463, 444

\bibitem[{Scargle (1982)}]{scargle82} Scargle, J. D. 1982, ApJ, 263, 835

\bibitem[{Simonetti et al. (1985)}]{simonetti85} Simonetti, J.~H., Cordes, J.~M., \& Heeschen, D.~S.\ 1985, ApJ,
296, 46

\bibitem[{Sitko et al. (1985)}]{sitko85} Sitko, M. L., Schmidt, G. D., \& Stein, W.
A. 1985, ApJS, 59, 323

\bibitem[{Skrutskie et al. (1997)}]{skrutskie97} Skrutskie, M. F., Schneider, S. E., Stiening, R.,
et. al.\ 1997, ASSL Vol.~210, Dordrecht: Kluwer, p. 25

\bibitem[{Stecker et al. (1996)}]{stecker96} Stecker, F. W., de Jager, O. C., \& Salamon, M. H. 1996, ApJ, 473, L75

\bibitem[{Tagliaferri et al. (2000)}]{tagliaferri00} Tagliaferri G., Ghisellini G., Giommi P. et al. 2000, A\&A, 354, 431

\bibitem[{Takalo (1991)}]{takalo91} Takalo, L. O. 1991, A\&AS, 90, 161

\bibitem[{Ter\"{a}sranta (2002)}]{terasranta02} Ter\"{a}sranta, H., 2002, in
Blazar Astroph. BeppoSAX \& Other Obs., ASI Special Publ., Rome
2002, p. 215

\bibitem[{Ter\"{a}sranta et al. (1998)}]{terasranta98} Ter\"{a}sranta, H., Tornikoski, M., Mujunen,
A., et al. 1998, A\&AS, 132, 305

\bibitem[{Thompson et al. (1995)}]{thompson95} Thompson, D.~J., Bertsch, D. L., Dingus, B. L.,  ~et
al.\ 1995, ApJS, 101, 259

\bibitem[{Tosti et al. (2002)}]{tosti02} Tosti, G., Massaro, E., Nesci, R., Ciprini, S., et al. 2002, A\&A, 395, 11

\bibitem[{Tosti et al. (1996)}]{tosti96} Tosti, G,, Pascolini, S., \& Fiorucci, M. 1996, PASP, 108, 706

\bibitem[{Valtaoja \& Ter\"{a}sranta (1996)}]{valtaoja96} Valtaoja, E.~\& Ter\"{a}sranta, H.\ 1996, A\&AS, 120, 491

\bibitem[{Valtaoja et al. (1993)}]{valtaoja93} Valtaoja, L., Karttunen, H., Efimov, Y. S., \& Shakhovskoy, N. M. 1993, A\&A, 278, 371

\bibitem[{Valtaoja et al. (1991)}]{valtaoja91} Valtaoja, L., Sillanp\"{a}\"{a}, A., Valtaoja, E., Shakhovskoi, N.
M., \& Efimov, I. S. 1991, AJ, 101, 78

\bibitem[{Wilkinson et al. (1998)}]{wilkinson98} Wilkinson, P. N., Browne, I. W. A., Patnaik, A. R., Wrobel, J.
M., \& Sorathia, B. 1998, MNRAS, 300, 790

\bibitem[{Wright et al. (1998)}]{wright98} Wright, S. C., McHardy, I. M., Abraham, R. G., et al. 1998, MNRAS, 296, 961

\bibitem[{Zacharias et al. (1999)}]{zacharias99} Zacharias, N., Zacharias, M. I., Hall, D. M. et al. 1999, AJ, 118, 2511

\end{thebibliography}

\label{lastpage}

\end{document}